\documentclass[pre,showpacs,twocolumn,amsmath,amssymb]{revtex4}
\usepackage{latexsym}
\usepackage{amscd}
\usepackage{amsthm}
\usepackage{epsf}
\usepackage{graphicx}    

\newcommand{\set}[1]{\left\lbrace #1 \right\rbrace} 
\newcommand{\defas}{\mathrel{\mathop{:}}=}   
\renewcommand{\subset}{\subseteq}  

\DeclareMathOperator*{\bigtimes}{\textnormal{\Large $\times$}} 
\DeclareMathOperator*{\supp}{supp}
\DeclareMathOperator*{\spann}{span}
\newcommand{\comm}[1]{ }
\providecommand{\abs}[1]{\left\lvert#1\right\rvert} 

\newcommand{\e}{\mathrm{e}} 
\newcommand{\ie}{i.e.\;}  

\newcommand{\iid}{i.i.d.\;} 

\theoremstyle{plain}
\newtheorem{thm}{Theorem}

\theoremstyle{definition}
\newtheorem{defn}[thm]{Definition}

\newtheorem{exmp}[thm]{Example}

\theoremstyle{remark}

\begin{document}

\title[Complexity from interaction]{Complexity Measures from
  Interaction Structures}

\author{T. Kahle$^{1}$, E. Olbrich$^{1}$, J. Jost$^{1,2}$, and
  N. Ay$^{1,2}$} \affiliation{$^{1}$ Max Planck Institute for
  Mathematics in the Sciences, Inselstrasse 22, D-04103 Leipzig,
  Germany}\email{kahle@mis.mpg.de} \affiliation{$^{2}$ Santa Fe
  Institute, 1399 Hyde Park Road, Santa Fe, NM 87501, USA}

\date{\today}
\begin{abstract}
  We evaluate information theoretic quantities that quantify
  complexity in terms of $k$-th order statistical dependencies that
  cannot be reduced to interactions among $k-1$ random
  variables. Using symbolic dynamics of coupled maps and cellular
  automata as model systems, we demonstrate that these measures are
  able to identify complex dynamical regimes.
\end{abstract}


\maketitle

\section{Introduction}
In this paper we study a notion of complexity based on interaction
among parts of a system. This is one of the first and most natural
choices and has been considered several times. Doing so, we will
pursue a geometric approach going back to the works of Amari
\cite{Amari01,Amari00}. In statistical physics, for example, the
complexity of a model is specified by the order of
interaction. Firstly, one considers the free theory with no
interaction, and as the next approximation a theory of pair
interaction. That means that there exists an energy function $H$,
assigning to each state of the world $x$ a real number $H(x)$ such
that the more probable states have a lower energy. If this function
has, using a suitable expansion into polynomials, only terms of order
two, one speaks of pair interaction. The probability to find the world
in the state $x$ at temperature $T$ is given by the Gibbs distribution
\begin{equation*}
P(x) = \frac{\e^{-\beta H(x)}}{Z(\beta)},
\end{equation*}
where, $k$ denoting Boltzmann's constant, $\beta$ is the inverse temperature
$\frac{1}{kT}$. By 
\begin{equation*}
Z(\beta) \defas \sum_{x} \e^{-\beta H(x)}
\end{equation*} 
we denoted the partition function, which is a normalization.

In this work, we study a complexity measure that incorporates the idea
to quantify the amount of pairwise, triple wise,\ldots\,-interaction
in a rigorous way. We do so by using the exponential families of
$k$-interactions which will be defined in Section \ref{sec:theory}. In
this context, a $k$-interaction is an interaction between $k$ random
variables. They can stand for $k$ different particles in the spatial
case, one variable in $k$ time steps, or a mixture of both.  Using the
notion of distance from such an exponential family, we can quantify
the amount of $k$-wise interaction that cannot be explained by
$(k-1)$-wise interaction. This allows us to quantify complexity. We
explore this concept using two examples: Symbolic Dynamics of Coupled
Map Lattices and Cellular Automata. For coupled map lattices we focus
on spatial interaction. Still, the temporal and spatiotemporal
features can be studied analogously as we will see for Cellular
Automata. Our approach quantifies all interactions in a system. It
does not necessarily refer to any neighborhood structure, any possible
subsystem can contribute to this interaction. Allowing every
subsystem, not necessarily local, to interact is motivated by possible
applications as for example the identification of positions in the
genome that carry a high information content.  Interacting regions in
the genome are not necessarily located spatially close to each
other. For example the arrangement of the chromosome in space can
bring sequentially distant sites into spatial proximity and thereby
enable regulatory interactions.

The paper is organized as follows. In Section \ref{sec:theory} we
introduce the necessary concepts from information theory and define
our model systems. We give details on the models and numerical methods
in Section \ref{sec:simulation} and in Section \ref{sec:results} we
give the results. Section \ref{sec:conclusion} contains concluding
remarks.

\section{Information Theory}
\label{sec:theory}
In this section we introduce exponential families of probability
measures, the Kullback-Leibler divergence and other notions.
Throughout the whole paper, the set of states of the system is of
compositional structure. That means, for each site index $v$ in a
finite set $V \defas \set{1,\ldots, N}$ we have a finite configuration
space $\mathcal{X}_v$. The set of all possible configurations is
\begin{equation*}
  \mathcal{X}_V \defas \bigtimes_{v\in V}\mathcal{X}_v,
\end{equation*} 
and likewise for each subset $A\subset V$, $\mathcal{X}_A \defas
\bigtimes_{v\in A}\mathcal{X}_v$.  Every real valued function on the
set $\mathcal{X}_V$ can be seen as an element of the vector space
\[
\mathbb{R}^{\mathcal{X}_V} \defas \set{f : \mathcal{X}_V \to
  \mathbb{R}}.
\]
As $\abs{\mathcal{X}_V}$ is finite, this space is isomorphic to
$\mathbb{R}^{n}$ with $n=\abs{\mathcal{X}_V}$. We will therefore call
elements in $\mathbb{R}^{\mathcal{X}_{V}}$ vectors or functions
without preference.  

\subsection{Families of Probability Measures} 
\label{sec:famil-prob-meas}
We now study probability distributions on the set
$\mathcal{X}_V$. Naturally, these are also elements of
$\mathbb{R}^{\mathcal{X}_V}$. Consider
\[
\overline{\mathcal{P}(\mathcal{X}_V)}
\defas \set{P \in \mathbb{R}^{\mathcal{X}_V} : P(x) \geq 0,\,
  \sum_{x\in\mathcal{X}_V} P(x) = 1},
\]  
the probability measures on the set $\mathcal{X}_V$. Due to the
compositional structure these probability measures are in fact only
joint probabilities of a set of random variables $\set{X_v : v\in V}$,
where $X_v$ takes values in $\mathcal{X}_v$.
$\overline{\mathcal{P}(\mathcal{X}_V)}$ has the geometrical structure
of a $(|\mathcal{X}_V| -1)$-dimensional simplex. For $P\in
\overline{\mathcal{P}(\mathcal{X}_V)}$, we call \[ \mbox{supp}(P)
\defas \set{x \in \mathcal{X}_V : P(x) > 0} \] the \emph{support} of
$P$. Then the probability distributions with full support are denoted
\[
\mathcal{P}(\mathcal{X}_V) \defas \set{P\in
\overline{\mathcal{P}(\mathcal{X}_V}) : P(x) > 0, \forall x\in\mathcal{X}_V}.
\]
The notation is justified as $\overline{\mathcal{P}(\mathcal{X}_{V})}$
is the closure of $\mathcal{P}(\mathcal{X}_{V})$.
\begin{exmp}
  We consider the example $V=\set{1,2}$ in the binary case, \ie
  $\mathcal{X}_1 = \mathcal{X}_2 = \set{0,1}$. We have $\mathcal{X}_V
  = \set{0,1}^{2} = \set{(00),(01),(10),(11)}$, and
  $\mathbb{R}^{\mathcal{X}_V}$ is just $\mathbb{R}^{4}$. The set of
  probability measures $\overline{\mathcal{P}(\mathcal{X}_V)}$ is a
  three-dimensional tetrahedron
  \begin{equation*}
    \set{(p_{00}, p_{01}, p_{10} , p_{11}) \in \mathbb{R}^4_{\geq0} 
      : p_{11} = 1 - p_{00} - p_{01} - p_{10}}.
  \end{equation*}
  It is the set of joint probabilities of two binary random
  variables. The extreme points of this tetrahedron are the $4$ unit
  vectors.
\end{exmp}

For two distributions $P,Q \in \overline{\mathcal{P}(\mathcal{X}_V)}$,
we can now define a notion of distance by
\begin{equation*}
  D(P\parallel Q ) \defas \left\lbrace 
  \begin{array}{ll}
    \sum_{x\in\mathcal{X}} P(x) \log_2
    \frac{P(x)}{Q(x)} & \mbox{ if } (*) \text{ holds}\\ 
    \infty & \mbox{ otherwise}.
  \end{array} \right.
\end{equation*}
Here, $(*)$ is the condition $\supp(P) \subset \supp(Q)$.
$D(P\parallel Q)$ is called the \emph{Kullback-Leibler divergence} or
\emph{relative entropy}. Although not a metric, it is non-negative and
equals zero if and only if $P \equiv Q$. 

In information geometry one studies families of probability
measures i.e. sub manifolds of $\overline{\mathcal{P}(\mathcal{X}_V)}$. A
very natural class of such families arises if we consider the
exponential map
\begin{equation*}
  \exp : \mathbb{R}^{\mathcal{X}_V} \to \mathcal{P}(\mathcal{X}_V) 
  \qquad f \mapsto \frac{\e^{f}}{\sum_{x\in\mathcal{X}_V} \e^{f(x)}}.
\end{equation*}
It acts by component-wise exponentiation and normalization.

\begin{defn}
  Let $\mathcal{I}$ be a linear subspace of $\mathbb{R}^{\mathcal{X}_V}$. We define the
  exponential family $\mathcal{E}_\mathcal{I}$ to be the image of $\mathcal{I}$
  under the exponential map:
  \[
  \mathcal{E}_{\mathcal{I}} \defas \exp (\mathcal{I}).
  \]
\end{defn}
Exponential families are well known in statistical science. They have many nice
properties with respect to maximum likelihood estimation and other inference
methods. For our application we are, for a given $P$, interested in minimizing
the distance $D(P \parallel Q)$ for $Q$ in a given exponential family. It can
be seen, that this is equivalent to maximum likelihood estimation of $P$ in the family 
$\mathcal{E}$.
To do so, let $\hat{P}$ be the relative frequencies of an observation. Then
the log-likelihood function with respect to $Q$ is $L(\hat{P},Q) \defas 
\sum_{x\in\mathcal{X}_V} \hat{P}(x) \log Q(x)$. Maximum likelihood estimation is to
find a $Q$ that maximizes this function. On the other hand, if we add to 
$-L$ the constant (with respect to $Q$) $\sum_{x\in\mathcal{X}_V}
\hat{P}(x)\log\hat{P}(x)$ we find 
\begin{align*}
- L(\hat{P},Q) & + \sum_{x\in\mathcal{X}_V}
\hat{P}(x)\log\hat{P}(x) \\  
 & = \sum_{x\in\mathcal{X}_V} \hat{P}(x)
\log{\frac{\hat{P}(x)}{Q(x)}} \\
& = D(\hat{P} \parallel Q)
\end{align*}
Therefore a maximizer of the likelihood is a minimizer of Kullback-Leibler
distance. 
On the other hand, geometrically, minimization of $D(P\parallel Q)$ could be
interpreted as a projection, since $D$ is a kind of distance. This view is
employed in information geometry. These projections are the key players in the
definition of our complexity measure.  

In what follows, we define the concrete exponential families that we
will use in the definition of our measure. The idea that we want to
follow was first explored by Darroch et al. \cite{darrochspeed80}. It
is to consider exponential families of $k$ interactions. Firstly, one
defines the linear space of functions depending on only $k$ of their
arguments. After taking the exponential map of sums of such functions,
we find probability distributions with only $k$-interactions. For a
given probability distribution, coming from a real system, we then
want to quantify its complexity by measuring how far it is from being
reducible to a theory of $k$-interactions.

\subsection{Interaction Spaces}
We exploit the compositional structure of $\mathcal{X}_V$ to 
define the notion of $k$-th order interactions. 

For any given $A\subset V$, we
write $x\in \mathcal{X}_V$ as $x = (x_A,x_{V\setminus
  A})$ \ie we distinguish between components in
  $A$ and outside $A$. Then we define ${\mathcal I}_A$ to be the
subspace of functions that do not depend on the configurations
outside $A$:
\begin{eqnarray*}
  {\mathcal I}_A \defas \left\{ f \in {\mathbb R}^\mathcal{X}_V : 
    f(x_A,x_{V \setminus A}) = f(x_A, x_{V
      \setminus A}') \right .\\ \left. \mbox{ for all $x_A \in \mathcal{X}_A$, $x_{V \setminus A}, 
      x_{V \setminus A}' \in \mathcal{X}_{V
        \setminus A}$} \right\}.
\end{eqnarray*}
Using these spaces as building blocks, one can define the interaction
space corresponding to interactions between $k$ arbitrary units
\[
   {\mathcal I}_k \defas \spann_{ A \subset V,\, |A| = k } {\mathcal I}_A,
\]
\ie just the span of all the vectors on the right hand side.
Associated with each of the interaction spaces is an exponential
family $\mathcal{E}_k \defas \exp(\mathcal{I}_k)$. The interaction
spaces are included in each other: $\mathcal{I}_1 \subset
\mathcal{I}_2 \subset \cdots \subset \mathcal{I}_N$, therefore we have
defined a hierarchy of exponential families:
\[ {\mathcal E}_{1} \; \subseteq \;
{\mathcal E}_{2} \subseteq \; \dots \subseteq \; {\mathcal E}_{N}.
\] 
  
This hierarchy was studied in \cite{Amari01,ayknauf06}. It has found various
applications in the theory of neural networks. The correspondence with the
notions of statistical physics is as follows: A vector $f \in \mathcal{I}_k$
corresponds to an energy which has only $k$-interactions, but no higher
interactions. It gives rise to a probability distribution 
\begin{equation*}
  P(x) = \frac{\e^{f(x)}}{Z} \in \mathcal{E}_k.
\end{equation*}
Vice versa every $P$ in, for example, $\mathcal{E}_2$ has a (non-unique) representation as 
\begin{align*}
  P(x) & = \prod_{A\subset V : \abs{A} = 2} \phi_A(x) \\
  & = \frac{1}{Z}\exp \left( \sum_{A\subset V: \abs{A} = 2} f_A(x_A) \right)
\end{align*} 
where the energy can be written with pair interactions only.

Note that $\mathcal{E}_N$, corresponding to all functions, is
$\mathcal{P}(\mathcal{X}_V)$. In general, in the image of the
exponential map, one has only distributions with full
support. Probability zero corresponds to infinite energy which is only
achievable by limits of sequences of probability measures in an
exponential family. It is an open question which of the possible
support sets are achievable by such limit probability measures in a
given exponential family. This question is connected with the face
structure of a convex polytope, the so called marginal polytope. See
\cite{CziMa02,KahleWenzelAy08} for details.  For our application one
can just pass to the closure(in $\mathbb{R}^n$) of the exponential
family. In practice, if $\hat{P}$ is the distribution given by the
relative frequencies observed in an experiment, we can compute the
minimizer of the Kullback-Leibler distance to an exponential family
using the iterative proportional fitting algorithm from
\cite{csiszarshields04}. However, the complexity of the algorithm
makes it unfeasible for values of $\abs{\mathcal{X}_V} \gg 10^6$.

Now we are about to introduce our complexity measures. They are based
on the notion of distance to an exponential family $\mathcal{E}$. We
define \[D(P \parallel \mathcal{E}) \defas \inf_{Q\in\mathcal{E}}
D(P\parallel Q).\] One can show that $D(P \parallel \mathcal{E})$ is
continuous with respect to $P\in
\overline{\mathcal{P}(\mathcal{X}_V)}$. Given two exponential families
$\mathcal{E} \subset \mathcal{F}$ on has 
\[
D(P \parallel \mathcal{E}) - D(P \parallel \mathcal{F}) \geq 0.
\]
We now use differences like these as components of a vector valued complexity
measure.

\subsection{Complexity measures}
\subsubsection{Definition}
For a given $P\in \overline{\mathcal{P}(\mathcal{X}_V)}$, we define a
vector valued complexity measure $I(P) \defas
(I^{(1)}(P),\ldots,I^{(N)}(P))$ with components
\begin{equation*}
  I^{(k)}(P) \defas D(P \parallel \mathcal{E}_{k-1}) - 
  D(P\parallel \mathcal{E}_k), \quad k=1,\ldots,N.
\end{equation*} 
In the next section we will argue for the choice and discuss
properties.

\subsubsection{Properties and Interpretation} 
The main idea is that $I^{(k)}$ quantifies those dependencies between
$k$ nodes that are not captured by interactions between smaller
subsets of nodes. However, this interpretation is a posteriori in the
following sense: Consider an open system, such as an infinite
CA. Although locally only pair interaction takes place, as the system
evolves in time we find higher order correlations. When we speak of
higher order interactions here, we mean an interpretation of higher
correlations by a model of higher order interaction in a closed
system.

We look at examples for our quantities. Consider first a distribution
$P$ which factors over the units, \ie it has a representation
\begin{equation*}
  P(x) = \prod_{v\in V} P_v(x_v),    
\end{equation*} 
where $P_v$ are the marginals on single units. In this case $P$ is an
element of the closure $\overline{\mathcal{E}_1}$ of the exponential
family $\mathcal{E}_1$ and therefore $I^{\left( k \right)}(P)$ equals
zero for $k \geq 2$.

For higher interactions a similar statement is true. One has that
$P\in\overline{\mathcal{E}_k}$ if $P$ admits a factorization as
\begin{equation}
  \label{eq:product_form}
  P(x) = \prod_{A : \abs{A} = k} \phi_A(x_A),
\end{equation} 
while here the converse is not true. A given $P \in
\overline{\mathcal{E}_{k}} \setminus \mathcal{E}_{k}$ need not have
this structure. In \eqref{eq:product_form} the functions $\phi_A$ have
the property to depend on their argument only through the part in $A$:
$x_A$. In this general case, the $\phi_{A}$ are not the marginals.  If
$P\in \overline{\mathcal{E}_{k}}$, then $I^{\left( l \right)}=0$ for
$l>k$.

Consider next a distribution in the synchronized case, \ie from the
value of one unit $v$ all the other units are determined. Take as an
example the distribution $P$ with $P(0\cdots0) = P(1\cdots1) =
\frac{1}{2}$ and $P(x) = 0$ for all other elements. It can be seen
that these distributions lie in the closure of $\mathcal{E}_2$, since,
loosely speaking, by pair interactions this behavior can be
explained. To do so let the second, third,\ldots\,unit copy the value
of the first unit, then the synchronized behavior is ``constructed''
using only conditions on pairs. Since $P \in \overline{\mathcal{E}_2}$
we have that $D(P\parallel \mathcal{E}_k) = D(P\parallel\mathcal{E}_2)
= 0$ for all $k>2$ and therefore the only non vanishing component of
$I(P)$ is $I^{\left( 2 \right)}$.

Another extreme case is given by the parity function. Let $P$ be the
uniform distribution on the following set of elements
\begin{equation*}
  \mathcal{Y} \defas \set{x : \mathcal{X}_V : x_1 = \sum_{i=2}^{N} x_i \mod 2}.
\end{equation*} $\mathcal{Y}$ is the set of
all configurations with even parity. It can be seen that for this distribution
$I^{\left( N \right)}$ is maximal while all other components vanish. This is due
to the fact, that every choice of $N-1$ units in is independent under $P$, while
there is total (functional) dependence between all the $N$ units. It can be seen
that even the reverse statement is true. We elaborate a bit on this. 
Let $g : \set{0,1}^{N-1} \to \set{0,1}$ be a binary function on the
$N-1$-strings. Define the following set 
\begin{equation*}
  \mathcal{Y}_{g} \defas \set{x\in \mathcal{X}_V : x = (g(y),y) \;
  y\in\set{0,1}^{N-1}}
\end{equation*} This means that $\mathcal{Y}_{g}$ is formed by those strings
where the nodes $2,\ldots,N$ are chosen freely and the first node has the value
of $g$. Let $P$ be the distribution that is uniform on the set
$\mathcal{Y}_g$. Then it can be shown that if $I^{N}(P)>0$ then either $g$ is the parity
function or $1 + g \mod 2$ is the parity function.

The vector $I$ has a natural correspondence to multi-information (or
integration as it is termed in
\cite{tononispornsedelman94}). Multi-information is a generalization
of mutual information and is defined as
\[
M(P) \defas \sum_{i\in V} H_i(P_i) - H(P),
\] 
where $H(P)\defas -\sum_{x\in\mathcal{X}_V} P(x)\log P(x)$ is the entropy of $P$
and $P_i$ is the $\set{i}$-marginal of $P$. Then $H_i(P_i) \defas
-\sum_{x\in\mathcal{X}_i}P_i(x)\log P_i(x)$ is the marginal entropy in the unit $i$.
Multi-information measures the distance of a distribution from independence of
the units. It allows a decomposition as 
\begin{equation*}
  M(P) = D(P \parallel \mathcal{E}_1) = \sum_{k=2}^{N} I^{(k)}(P).
\end{equation*} 
See \cite{jostaybertschiolbrich06} for these complexity measures,
their properties, and the relation with TSE Complexity
\cite{tononispornsedelman94}.

In \cite{ayknauf06}, it is shown that the maximizer of
multi-information are contained in the closure of the exponential
family $\mathcal{E}_2$ and therefore have the $I$-vector concentrated
in $I^{(2)}$.

In the following we compare our approach with the one of
\cite{Lindgren1987}. There the author distinguishes a chemical
contrast, which corresponds to our $I^{\left( 1 \right)}$. It
quantifies the information that is contained in a sequence due to the
average density of zeros differing from $\frac{1}{2}$. Then,
quantities $k_m$, the correlational contrasts, were defined. These are
in fact conditional mutual information, which measure the difference
of the actual distribution to a Markov approximation. They also appear
in the effective measure complexity \cite{Grassberger86d} which can be
expressed as a weighted sum over these quantities.  The relation to
our measures is that $k_m=0$ implies $I^{\left( m \right)}=0$. This
can be understood from the fact that a distribution that has a Markov
property with respect to $m-1$ marginals will be representable by
$m-1$ interactions. On the other hand, a distribution that is
representable by pair interactions does not necessarily admit a Markov
representation as the example
\begin{equation*}
  P(x)=\frac{1}{Z}\exp\left\{ c_1x_1x_2 + c_2 x_2x_3 + c_3 x_3x_1 \right\}
\end{equation*} of a triangle with no three-way interaction shows.
Therefore our measures give a finer view onto the interaction structure.

Having identified ``non complex'' behaviors with extreme values of the
components of $I\left( P \right)$ we now look for dynamical systems
where $I$ is not concentrated in one component, because if several
$I^{(k)}$ contribute, interactions on different scales coexist. This
then could indicate complex behavior.  We will next introduce our
model systems where we try to find complex behavior in the above
sense.

\section{Models and Methods}
\label{sec:simulation} 
\subsection{Model Systems}

\subsubsection{Cellular Automata}
A possible natural choice for a model system are one dimensional
(binary) cellular automata. They produce discrete output and are
therefore ideally suited for our method. Furthermore, CA were
introduced by S. Wolfram\cite{ wolfram1983} as paradigmatic models to
study complexity.  In this work, we restrict ourselves to some of the
256 elementary binary CA studied by Wolfram \cite{wolfram94}. For
these, starting from a configuration $x^{(t)} \in \set{0,1}^{N}$ the
next configuration $x^{\left( t+1 \right)}$ at a site $i$ is computed
as
\begin{equation*}
  x^{\left( t+1 \right)}_i = f(x^{(t)}_{i-1},x^{(t)}_i,x^{(t)}_{i+1}) 
\end{equation*}
for some function $f : \set{0,1}^{3}\to\set{0,1}$. Typically one
chooses periodic boundary conditions, \ie $x_{N+1}$ is identified with
$x_1$. There exist exactly 256 functions $\set{0,1}^{3}\to\set{0,1}$,
each characterized by its binary vector of values in
$\set{0,1}^{8}$. These vectors correspond to binary representations of
integers smaller than 256 and the rules get numbered accordingly.

In this simple model, only the next neighbors of node $i$ influence
this particular node. A priori, there are 88 elementary CAs modulo the
inversion and reflection symmetry. Furthermore, there are trivial
cases like rule 0, etc. such that the number of ``interesting'' rules
is smaller. In this paper, we pick out some specific rules that have
found attention in the literature.  Specifically, we study the rules
18, 20, 22, 30, 45, 50, 90, 110, 126 and 150.

\subsubsection{Coupled Map Lattices}
\label{sec:coupledmapmodel}
Our second model system: Symbolic Dynamics of Coupled
Map Lattices. Let again $V=\set{1,\ldots,N}$ denote the set of node indexes. Assume we are
given a connection structure between the nodes, specified by a graph $G$ with
vertices $V(G) = V$. By abuse of notation, the structure of the
network is given by the symmetric adjacency matrix $G =
(g_{ij})_{i,j=1,\ldots,N}\in \set{0,1}^{N\times N}$ of the 
graph. We study the discrete time case $t=0,1,2,\ldots$ and every node
$i\in V$ carries a real value $x_i(t) \in [0,1]$. These values get
updated simultaneously according to the so called coupled tent map
rule \cite{rev-kaneko}
\begin{equation*}
  x_i(t+1) = \epsilon \sum_{j} \frac{1}{k_i} g_{ij} f(x_j(t)) + 
  (1-\epsilon) f(x_i(t)),
\end{equation*}
where $k_i$ is the number of neighbors of node $v$ and $f: [0,1] \to
[0,1]$ defined by
\begin{equation}
  \label{eq:Coupled_Tent_Map_Rule}
  x \mapsto f(x) = \left\lbrace
\begin{array}{ll}
  2x  & \mbox{ if } 0\leq x \leq \frac{1}{2} \\
  2(1-x)  & \mbox{ if } \frac{1}{2}\leq x \leq 1 
\end{array} \right. .
\end{equation} 
is the tent map on the unit interval. Tent map lattices are a well
established model to study synchronization and pattern formation in
spatially extended systems \cite{couptent-syn,couptent-macky03}.

We will study binary symbolic dynamics of coupled tent map network
\cite{book-symbolic, jalanjostatay-chaos06}. For each node we consider
$\mathcal{X}_i \defas \set{0,1}$ and for every $A\subset V$ we have
$\mathcal{X}_A = \set{0,1}^{\abs{A}}$ as in Section \ref{sec:theory}.
Then the configuration space of the symbolic dynamics is
$\mathcal{X}_V = \set{0,1}^N$, the space of binary sequences of length
N. To every real valued dynamics $x_i(t)$ of node $i$, we consider the
symbolic dynamics
\begin{equation*}
  s_i(t) = \left\lbrace
  \begin{array}{ll}
    0 & \mbox{ if } 0 \leq x_i(t) \leq x^* \\
    1 & \mbox{ if } x^* < x_i(t) \leq 1\\
  \end{array} \right.
\end{equation*}
where $x^* \in [0,1]$ is a given value. We
obtain a global configuration of symbols $s(t) =
(s_1(t),\ldots,s_{N}(t)) \in \mathcal{X}_V$. Our aim is to utilize
this symbolic time-series in order to evaluate the complexity of the
full dynamics.  Assuming that a dynamical system has arrived in a
stationary state it will produce a stationary distribution on the set
of symbolic strings. The method we propose is to capture this
stationary distribution $P$ and evaluate the complexity $I(P)$.  We
will call a dynamics complex if it exhibits interactions of different
orders at the same time.

\begin{figure*}[htpb]
  \centering
   \resizebox{0.95\textwidth}{!}{\includegraphics[angle=-90]{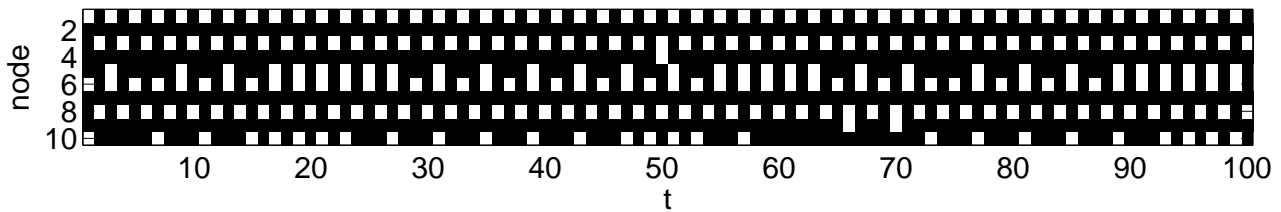}}
   \caption{\label{fig:snapshotdynamicscirc} A snapshot of the highly
     regular binary symbolic dynamics of the circle graph at
     $\epsilon=0.476$. Each row shows one node over time. White spots
     indicate the value zero, black squares one.}
\end{figure*}

\begin{figure*}[htpb]
  \centering
   \resizebox{0.95\textwidth}{!}{\includegraphics[angle=-90]{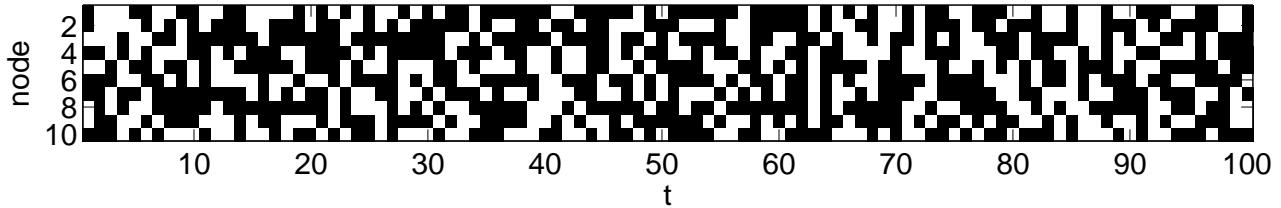}}
   \caption{\label{fig:snapshotchaotic} A snapshot of the weakly
     coupled chaotic dynamic of the circle graph at
     $\epsilon=0.04$. Each row shows one node over time. White spots
     indicate the value zero, black squares one.}
\end{figure*}

\begin{figure*}[htpb]
  \centering
  \resizebox{0.95\textwidth}{!}{\includegraphics[angle=-90]{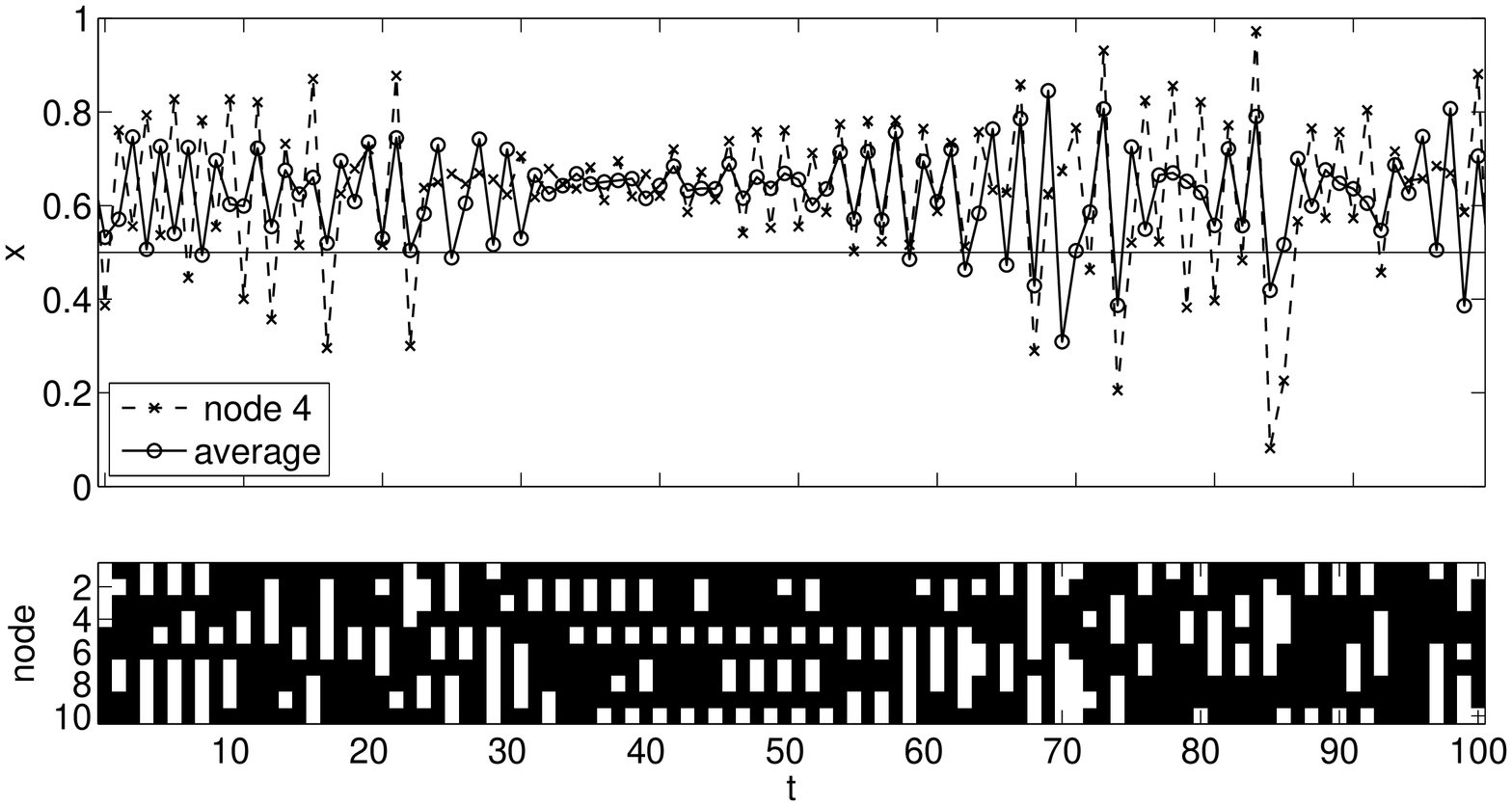}}
  \caption{\label{fig:snapshotcomplex} A snapshot of the complex
    dynamic of the circle graph at $\epsilon=0.482$. In the top the
    activity of node 4 is plotted together with the average activity
    over all nodes. One can observe that for short periods of time
    synchronization effects occur, for instance between $t=30$ and
    $t=40$ node 4 is almost constant while the average also fluctuates
    less. Below the symbolic dynamics, with black as one, white as
    zero, of all 10 nodes is shown. The synchronization is visible
    here too as the pattern after $t=35$ is very regular.}
\end{figure*}

\subsection{Simulation and Computation}
\subsubsection{Cellular Automata}
As our computational power limits us to analyzing short sequences, we
cut pieces of length 14 from CA of length 20000.  These were iterated
for $10^{6}$ time steps, starting from an \iid initial condition. To
gain more insight in the structure of our measures we furthermore
evaluate it for the ``interaction cone'' corresponding to a single
time step. To do so we take all binary strings of length 3 and compute
the value of the CA on these strings. We then append the output as the
4th digit to the input string and compute $I^{\left( 1
  \right)},\ldots,I^{(4)}$ on the uniform distribution on these
strings. Of course in the running CA one has to consider the influence
of the stationary distribution on the inputs. To measure this
influence we also evaluate the elementary 4-strings when the input is
sampled with the stationary distribution.
\subsubsection{Coupled Map Lattice}
To simulate the coupled map lattice one has to specify a graph. For our
simulations, we use a fully connected graph(``globally coupled
network''), and a circle graph (``next neighbor coupling'') of 10
nodes each. We initialize the networks by setting the $x_i(0)$ to
independently uniformly distributed random values in the real
interval $(0,1)$.  The dynamics generates two time-series, the real
valued one $x(t)=(x_1(t),\ldots,x_{N}(t))$ and, using
$x^*=\frac{1}{2}$, the symbolic one $s(t) =\left( s_1(t),\ldots,
  s_N(t)\right)$.  In the uncoupled case, the value $x^* =
\frac{1}{2}$ gives the generating partition. It makes the symbolic
time series most informative about the true time series. We discard
the symbols during a transient time, assuming that after the transient
time (chosen here as $10^{6}$ time steps), the system is in a
stationary state.  Treating the symbolic time-series as $N$
realizations of a random variable with probability distribution $P \in
\overline{\mathcal{P}(\mathcal{X}_V)}$ on the space of symbols we
compute $I(P)$ for this $P$.

We also experimented with different values of the bound $x^*$,
but it turned out that this does not change the qualitative behavior. The
value $x^{*} = \frac{2}{3}$ has received some attention as
synchronization of the real valued dynamics can be detected from the
symbols \cite{jalanjostatay}. For our measures however, $I$ does not
seem to depend on the choice of $x^{*}$ in such an essential manner.

\subsection{Computation of $I^{(k)}$.}
\label{sec:computation-ik}
As we have seen in Section \ref{sec:famil-prob-meas}, computing an
information projections, \ie a minimizer of $D(P\parallel Q)$, for $P$
given and $Q \in \overline{\mathcal{E}}$ can be written as computing a
maximum likelihood estimation in some discrete statistical model. This
estimation can be solved analytically only for some very simple
models, such as $\mathcal{E}_{1}$. Nevertheless, for the general case
good iterative algorithms exist. These go back to Kullback
\cite{kullback68:_infor_theor_and_statis,csiszar75:_i_diver_geomet_of_probab}.
We have reimplemented the algorithm in parallel in the program
\texttt{CIPI} which is freely available \cite{cipi}.  As the
projections can be computed, the determination of $I^{(k)}$ is merely
a subtraction of two KL-divergences.

\section{Results}
\label{sec:results}

\subsection{Cellular Automata}
In Table \ref{tab:cellautres} the experimental values for sequences of
length 14, cut from large cellular automata are given. Then in Tables
\ref{tab:cellaut31uniform} and \ref{tab:cellaut31specific} the values
of the $I$-Vector for elementary steps of each of the automata are
printed. We distinguish between the uniform input and the true input
statistic sampled from the stationary distribution of the respective
CA.

\begin{table}
  \centering
  \begin{tabular}{|c||c|c|c|c|c|c|c|c|}
    \hline
    rule & $I^{(1)}$ & $I^{(2)}$ & $I^{(3)}$ & $I^{(4)}$ & $I^{(5)}$ & $I^{(6)}$
    & $I^{(7)}$ & $I^{(8)}$ \\ 
    \hline
    \hline
    18 & 1.828 & 2.237 & 0.031 & 0.010 & 0.002 & 0 & 0 & 0\\
    \hline
    20 & 1.890 & 1.556 & 0.256 & 0.028 & 0.003 & 0 & 0 & 0\\
    \hline
    22 & 0.634 & 0.479 & 0.076 & 0.179 & 0.077 & 0.035 & 0.019 & 0.006 \\
    \hline
    30 & 0 & 0 & 0 & 0 & 0 & 0 & 0 & 0 \\
    \hline
    45 & 0 & 0 & 0 & 0 & 0 & 0 & 0 & 0 \\
    \hline
    50 & 0 & 9.011 & 0 & 0 & 0 & 0 & 0 & 0 \\
    \hline
    54 & 0.007 & 2.658 & 0.181 & 1.409 & 0.040 & 0.004 & 0 & 0 \\
    \hline
    90 & 0 & 0 & 0 & 0 & 0 & 0.002 & 0.002 & 0.002 \\
    \hline
    110 & 0.156 & 3.738 & 2.070 & 0.035 & 0.009 & 0 & 0 & 0 \\
    \hline
    126 & 0 & 2.809 & 0.307 & 0.095 & 0.004 & 0 & 0 & 0 \\
    \hline
    150 & 0 & 0 & 0 & 0 & 0 & 0 & 0 & 0 \\
    \hline
  \end{tabular}
  \caption{Results for Cellular Automata, $0$ stands for a value less than
  $10^{-3}$. $I^{(9)}$ to $I^{\left( 14 \right)}$ are not shown, as they vanish.}
  \label{tab:cellautres}
\end{table}

\begin{table}
  \centering
  \begin{tabular}{|c||c|c|c|c|}
    \hline
    rule & $I^{(1)}$ & $I^{(2)}$ & $I^{(3)}$ & $I^{(4)}$ \\ 
    \hline
    \hline
    18 & 0.189 & 0.311 & 0.500 & 0  \\
    \hline
    20 & 0.189 & 0.311 & 0.500 & 0 \\
    \hline
    22 & 0.046 & 0.158 & 0.800 & 0  \\
    \hline
    30 & 0 & 0.189 & 0.811 & 0 \\
    \hline
    45 & 0 & 0.189 & 0.811 & 0  \\
    \hline
    50 & 0.046 & 0.885  & 0.069 & 0  \\
    \hline
    54 & 0 & 0.189 & 0.811 & 0 \\
    \hline
    90 & 0 & 0 & 1 & 0 \\
    \hline
    110 & 0.046 & 0.158 & 0.796 & 0 \\
    \hline
    126 & 0.189 & 0 & 0.811 & 0 \\
    \hline
    150 & 0 & 0 & 0 & 1 \\
    \hline
  \end{tabular}
  \caption{Cellular Automate $3\to 1$ statistics with uniform inputs.}
  \label{tab:cellaut31uniform}
\end{table}

\begin{table}
  \centering
  \begin{tabular}{|c||c|c|c|c|}
    \hline
    rule & $I^{(1)}$ & $I^{(2)}$ & $I^{(3)}$ & $I^{(4)}$ \\ 
    \hline
    \hline
    18 & 0.757 & 0.451 & 0.635 & 0  \\
    \hline
    20 & 0.822 & 0.403 & 0.535 & 0 \\
    \hline
    22 &0.260 & 0.135 & 0.908 & 0  \\
    \hline
    30 & 0 & 0.189 & 0.811 & 0 \\
    \hline
    45 & 0 & 0.188 & 0.812 & 0  \\
    \hline
    50 & 0 & 3 & 0 & 0  \\
    \hline
    54 & 0.003 & 0.160 & 0.847 & 0 \\
    \hline
    90 & 0 & 0 & 1 & 0 \\
    \hline
    110 & 0.059 & 0.241 & 0.804 & 0 \\
    \hline
    126 & 0 & 0.500 & 1 & 0 \\
    \hline
    150 & 0 & 0 & 0 & 1 \\
    \hline
  \end{tabular}
  \caption{Cellular Automate $3\to 1$ statistics with specific 
    input statistics.}
  \label{tab:cellaut31specific}
\end{table}

The values in Table \ref{tab:cellautres} represent properties of the
invariant measure of the respective CA. We confirm, that rules 30,45,
and 150 have the uniform distribution as their invariant measure,
therefore no interaction can be detected. The results for rule 22 show, that the long
range correlations, known from \cite{grassberger86b, grassberger1999},
can be detected using our measure.

In the elementary interaction cone of $3+1$ elements, we can observe
the following principles: Rule 18 and rule 20 build up correlations in
the input which leads to overall stronger correlations when the
stationary distribution is used. One can also see, that the outputs
are not uniformly distributed leading to a high value of $I^{(1)}$
when the inputs are sampled from the stationary distribution.  In the
elementary time step of rule 22 a lot of triple wise interactions are
detected.  This property is also found when the inputs are sampled
from the stationary distribution. Rule 50 is a very simple rule which
generates a periodic pattern of period 2. We see this represented here
as (like in the synchronized case, described in the introduction) the
$I$-vector is concentrated in $I_2$. When the inputs are uniform on
the other hand, this rule has some small correlations of three
positions, these get eliminated as the system reaches the stationary
distribution.  Rule 90 and rule 150 are additive rules that are
modeled by the XOR function.  Rule 150 takes XOR of every input,
giving exactly the correlation between all 4 digits that was described
in Section \ref{sec:theory}. Rule 90 is similar, it computes XOR of
the inputs 1 and 3, leading to only triple wise interaction.

For rules 22,30,45 there is evidence for the long range correlations
in \cite{grassberger1987}. Here, we can distinguish rule 22 from the
other two, as rule 30 and rule 45 still show the uniform distribution
in the outputs leading to $I^{(1)}$ being zero. Rule 30, Stephen
Wolfram's ``all time favorite rule'', is believed to have a high
degree of randomness. (It might actually be used as a random number
generator in \texttt{Mathematica}). For us it is not distinguishable
from rule 45.  Rule 110, which is capable of universal computation
shows no specific behavior with respect to our measure.  Rule 126 has
a mirror symmetry leading to $I^{\left( 2 \right)}$ being zero when
the input is sampled uniformly. However, the stationary distribution
induces a different symmetry. This leads to an interesting statistics,
which looks as follows
\begin{equation*}
  p_{126}(x) =\begin{cases}
    \frac{1}{8} & \text{ for $x \in \set{(001),(011),(100),(110)}$} \\
    \frac{1}{4} & \text{ for $x \in \set{(000),(111)}$} \\
    0 & \text{ otherwise.}
  \end{cases}
\end{equation*}
It can be seen that the rule exactly computes XOR of input 1 and 3 on
these configurations (which it does not do on all
configurations). This leads to the contribution of $I^{(3)}$. On the
other hand, the nonexistence of the configurations $(101)$ and $(010)$
can only be accomplished by additional interactions of order 2. A
comparison with rule 90 shows the non-additive nature of $I$. The
exactness of the probabilities accounts for the exact values in
$I$. In fact these values can be predicted from theory.

\subsection{Coupled Tent Maps}
\subsubsection{Fully connected Graph}

\begin{figure}[htp]
  \centering
  \includegraphics[height=4.8cm]{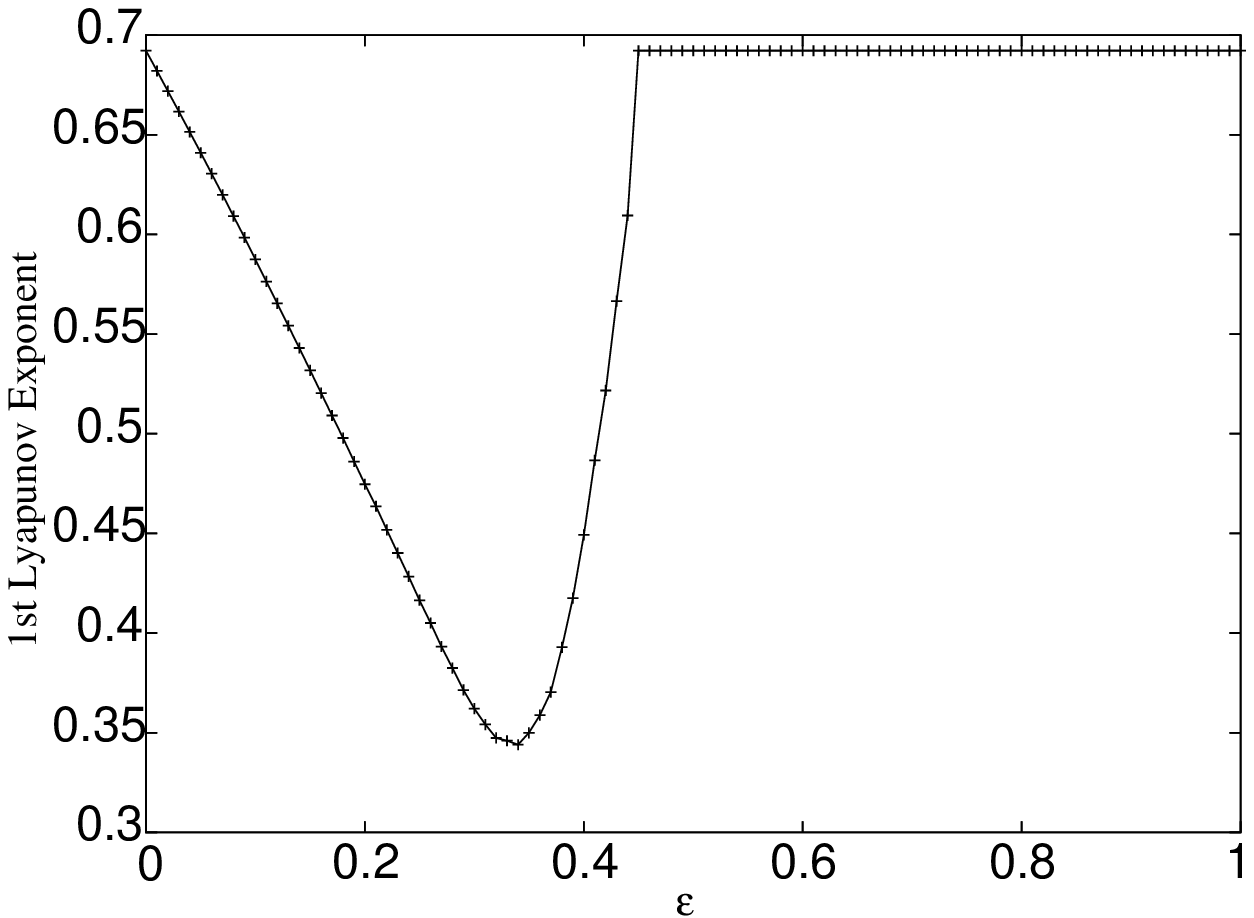}
  \includegraphics[height=4.8cm]{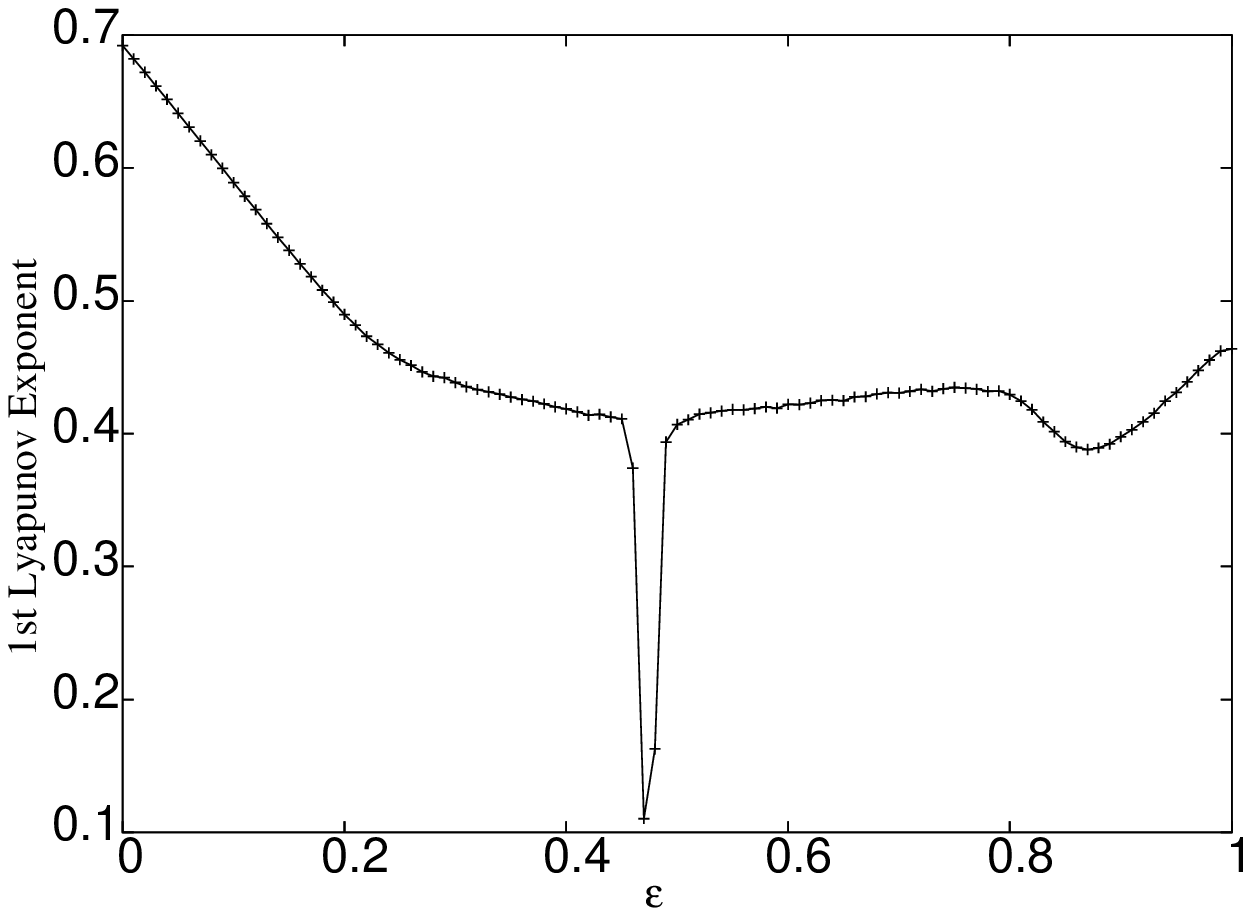}
  \caption{Behavior of the largest Lyapunov exponent $\lambda_{1}$ for
    the fully connected graph(top), and the circle graph(bottom).
    Note that the minima are attained in exactly the same parameter
    region, in which the $I^{(k)}$ have maxima.}
  \label{fig:lyapu}
\end{figure}

\begin{figure*}[pht]
  \centering
  \includegraphics[height=6.35cm]{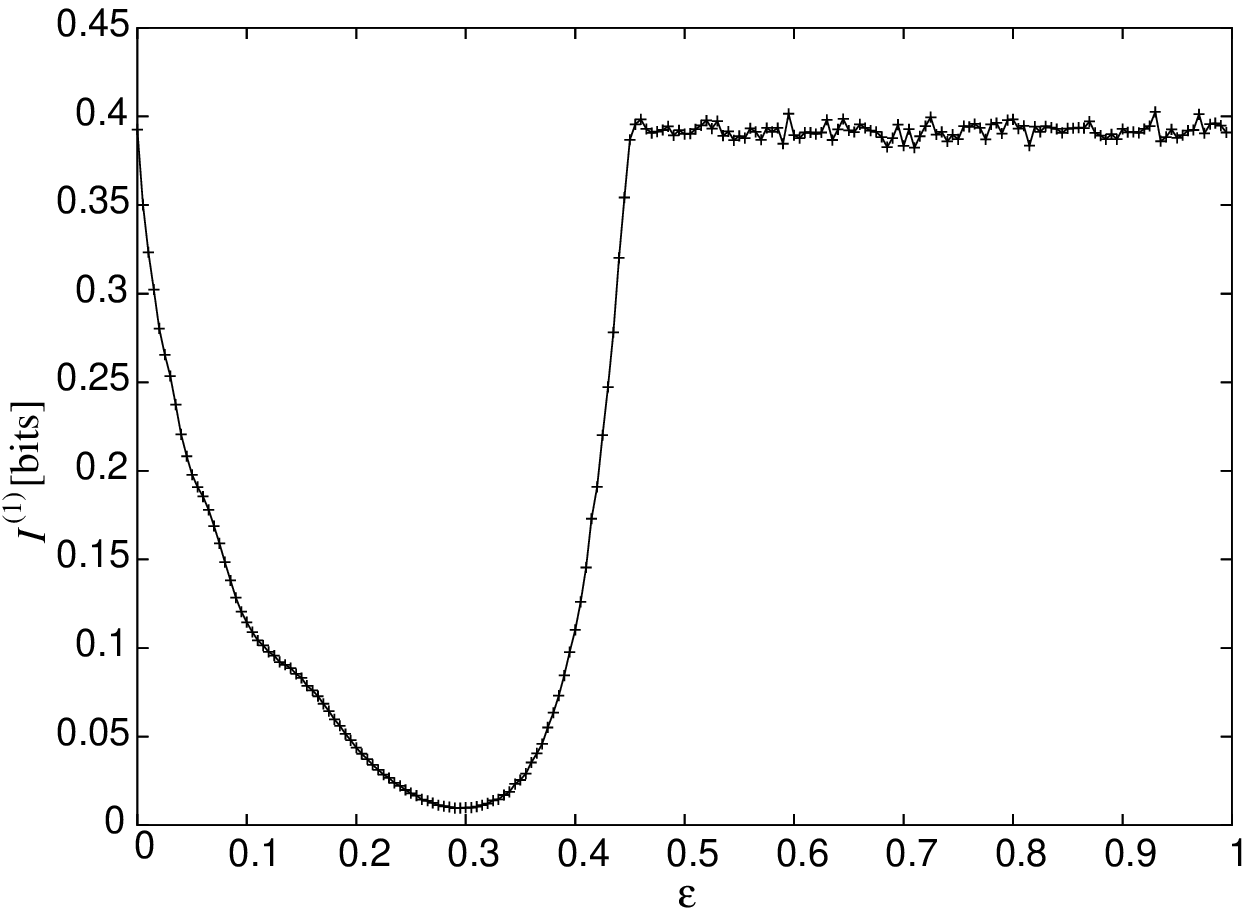}
  \includegraphics[height=6.35cm]{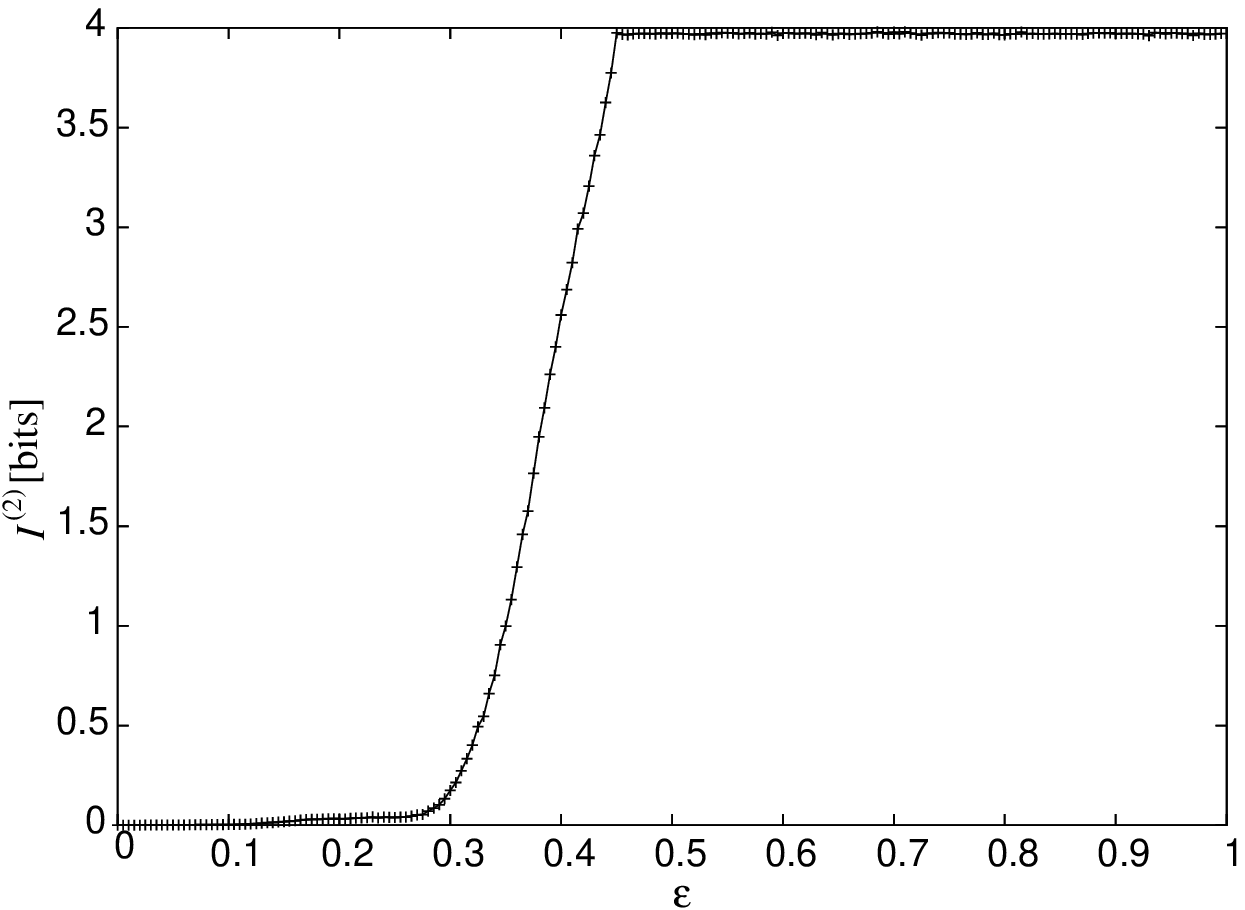}
  \includegraphics[height=6.35cm]{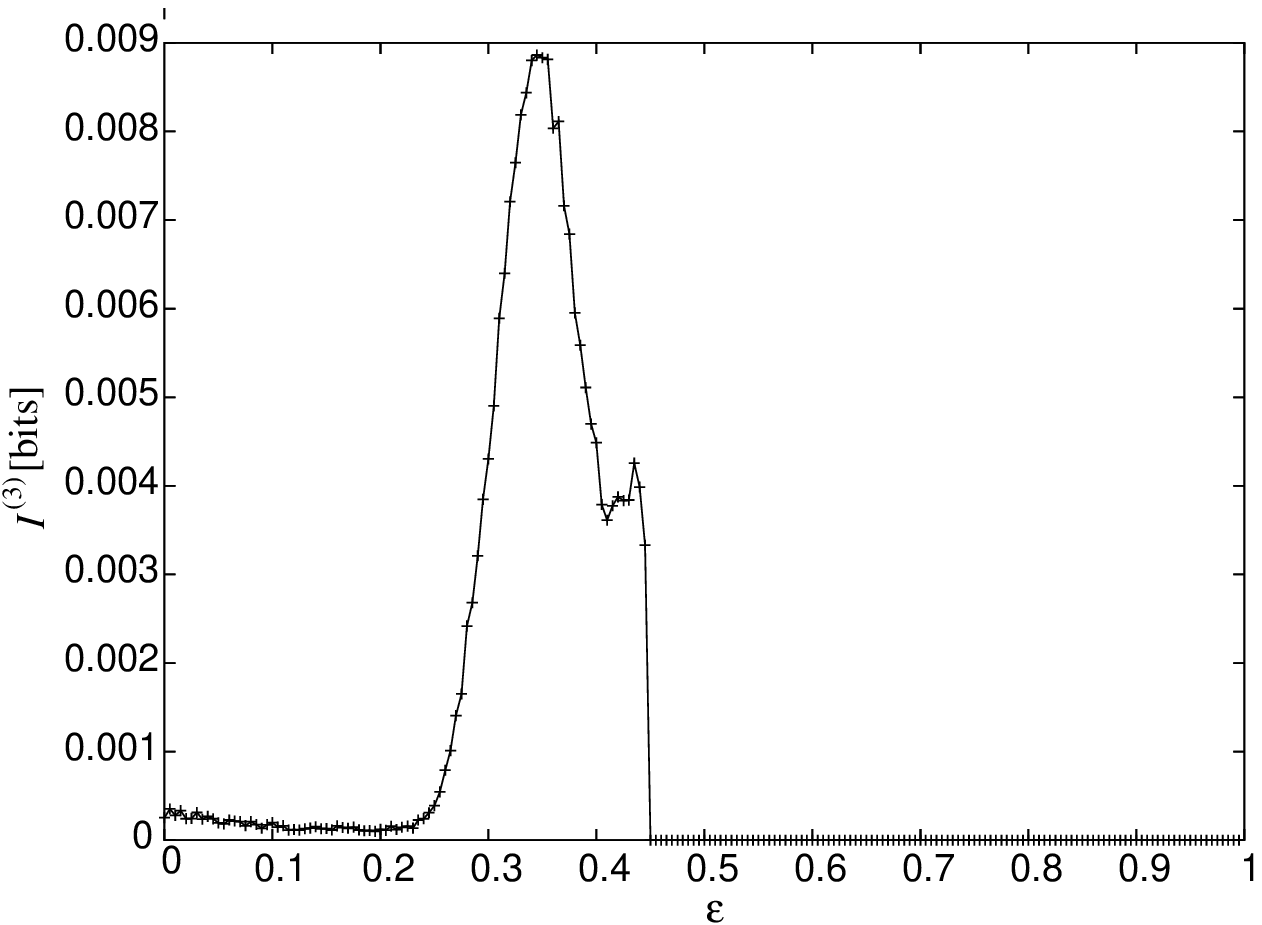}
  \includegraphics[height=6.35cm]{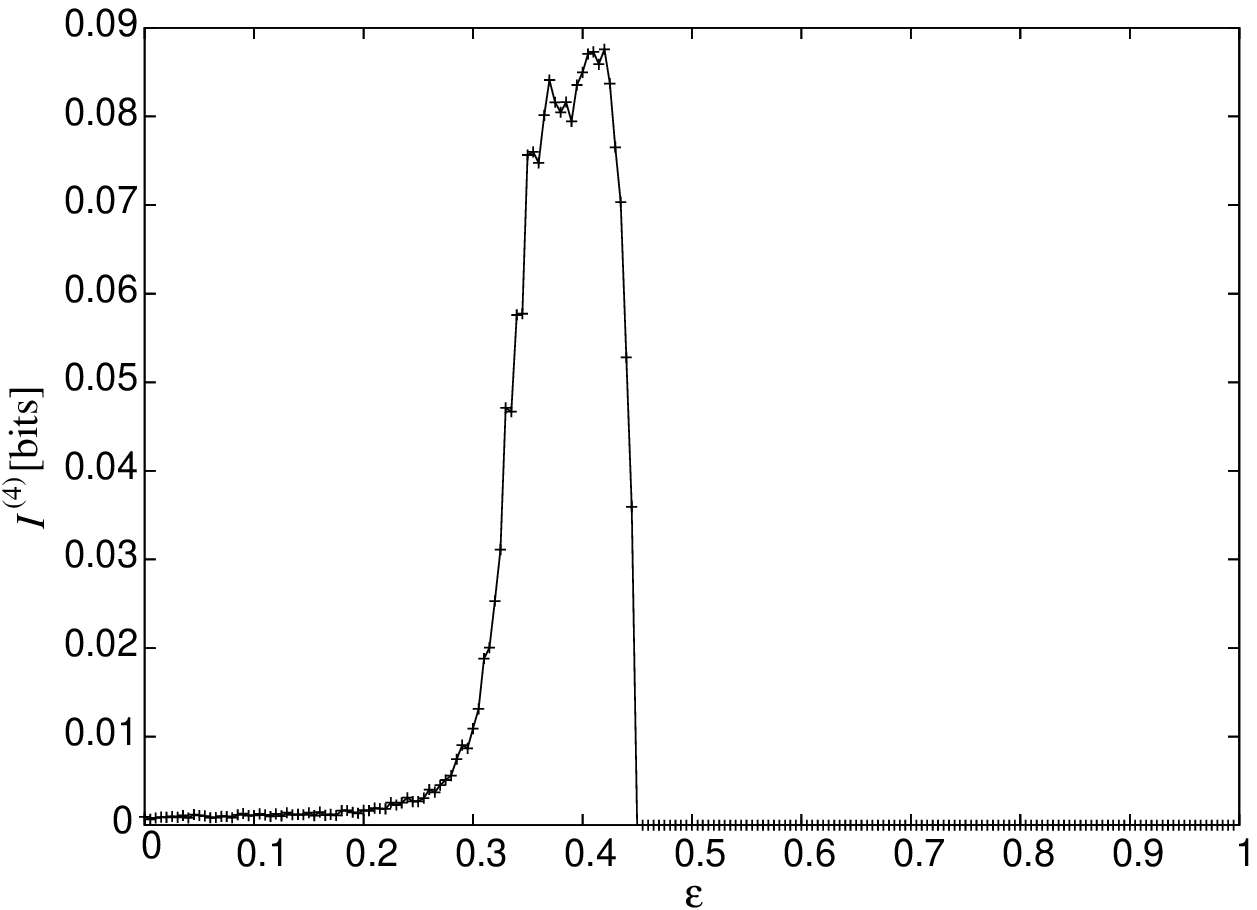}
  \includegraphics[height=6.35cm]{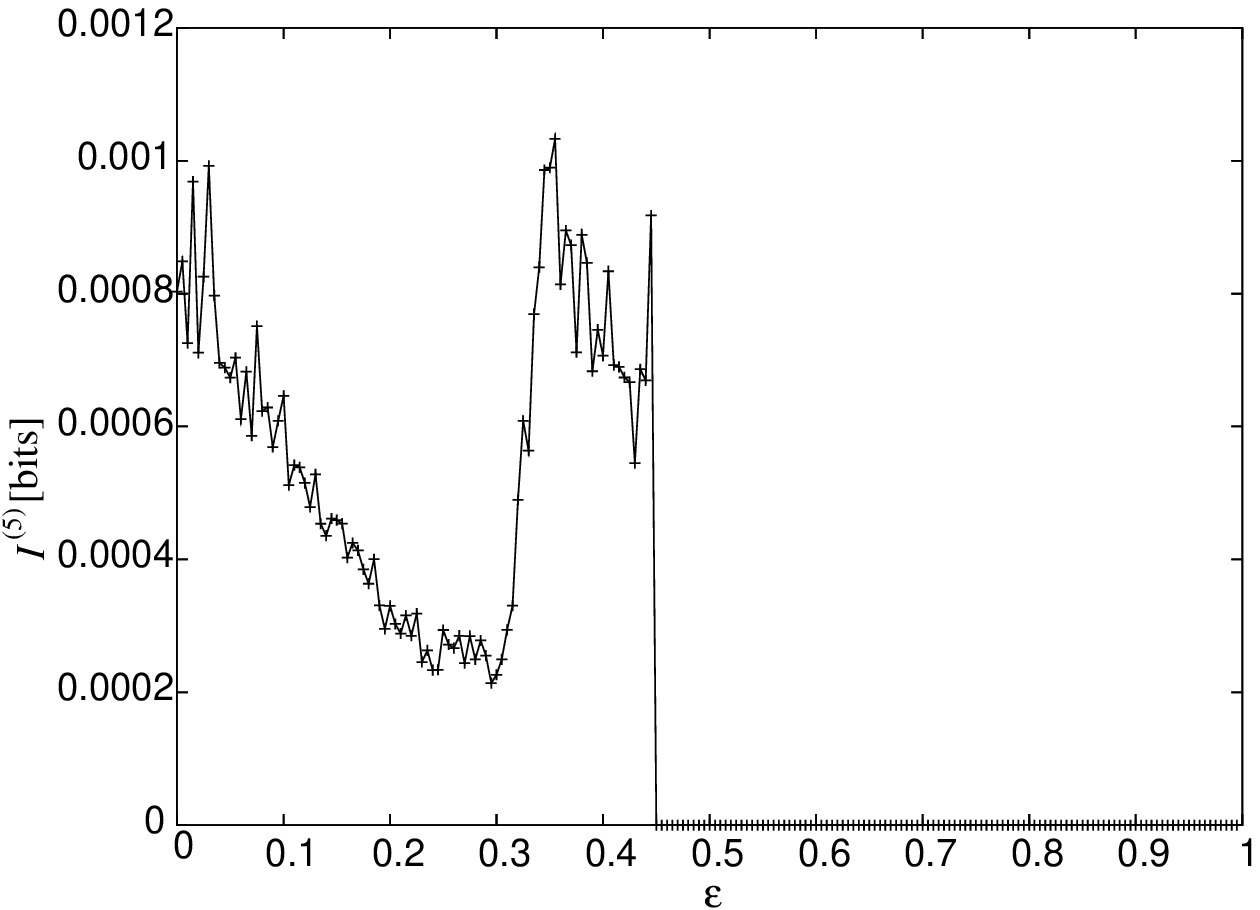}
  \includegraphics[height=6.35cm]{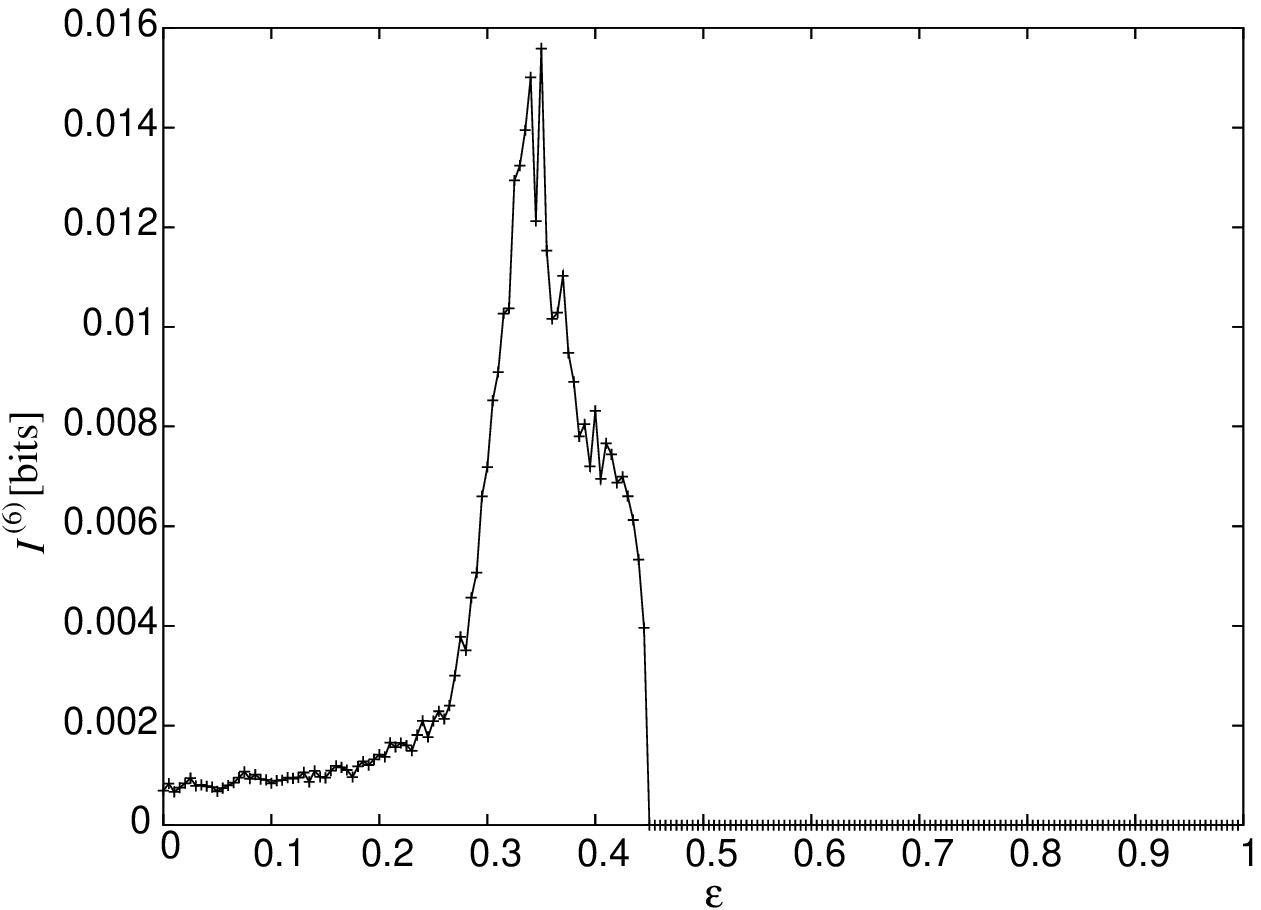}
  \caption{\label{fig:full} The Values of $I^{(k)}$ for the fully
    connected graph. For $\epsilon > 0.45$, the network is in the
    state of synchronized chaos. This is fully captured by pair
    interactions, therefore $I^{(2)}$ is maximal, while the $I^{(k)}$
    for $k>2$ vanish. On the edge to synchronized chaos high
    correlations on different scales are detected.}
\end{figure*}

\begin{figure*}[pht]
  \centering
  \includegraphics[height=6.35cm]{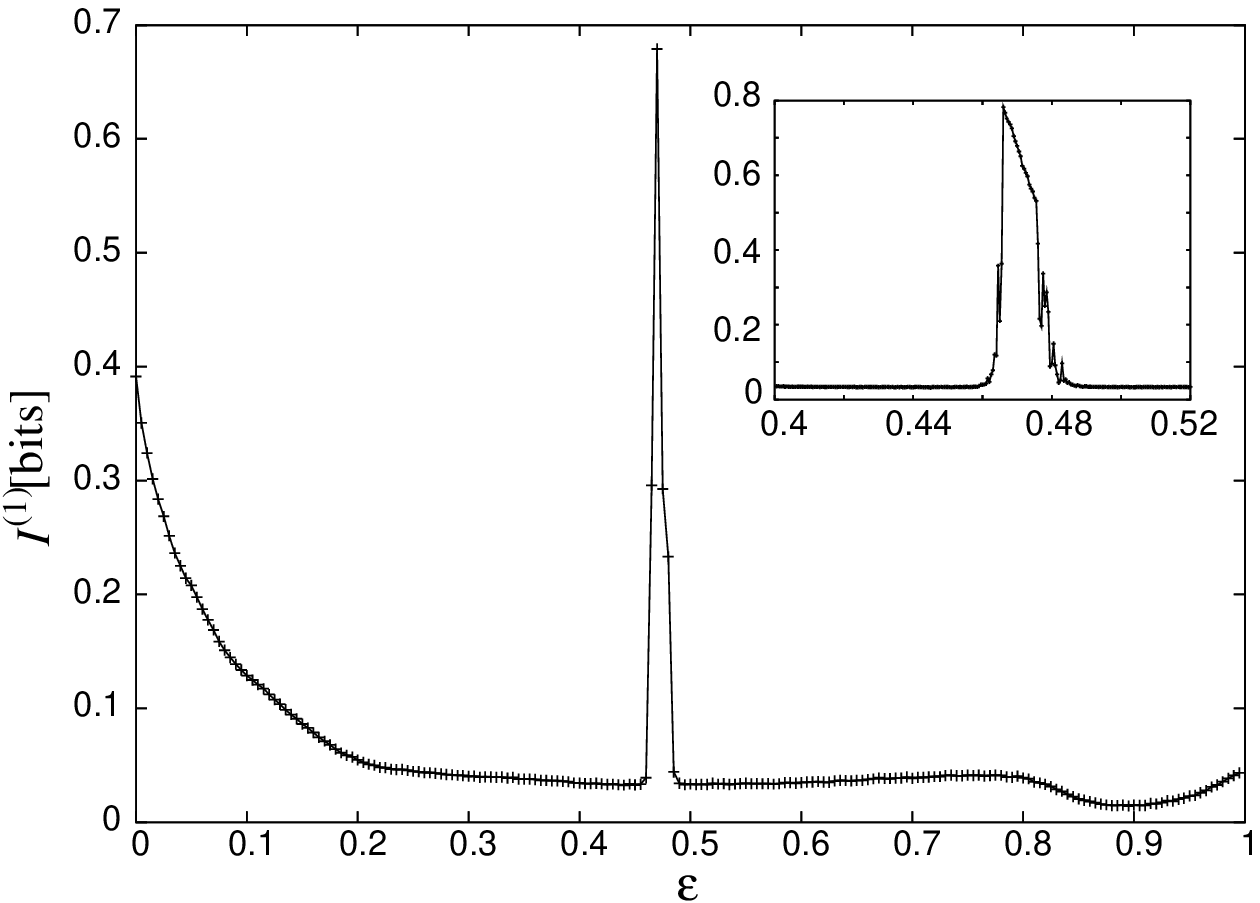}
  \includegraphics[height=6.35cm]{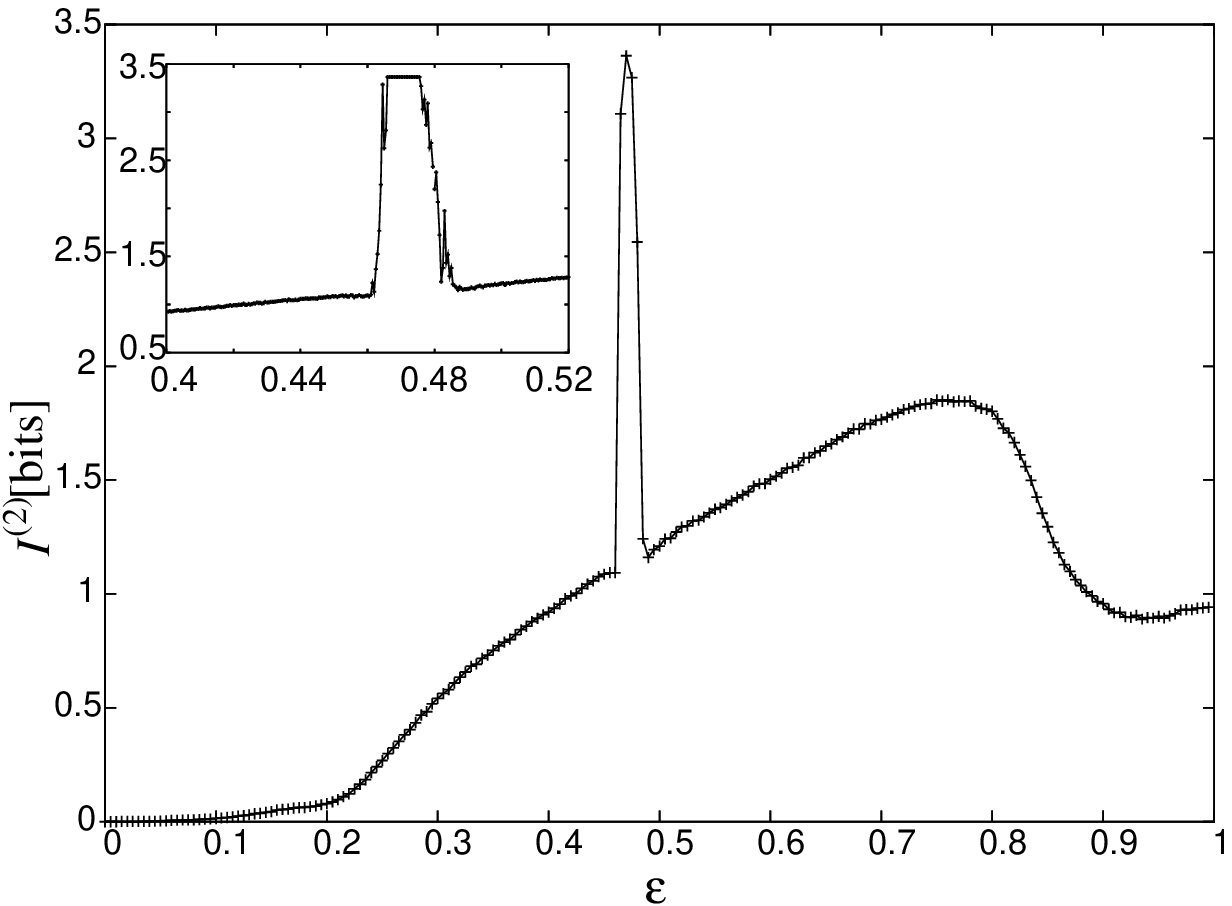}
  \includegraphics[height=6.35cm]{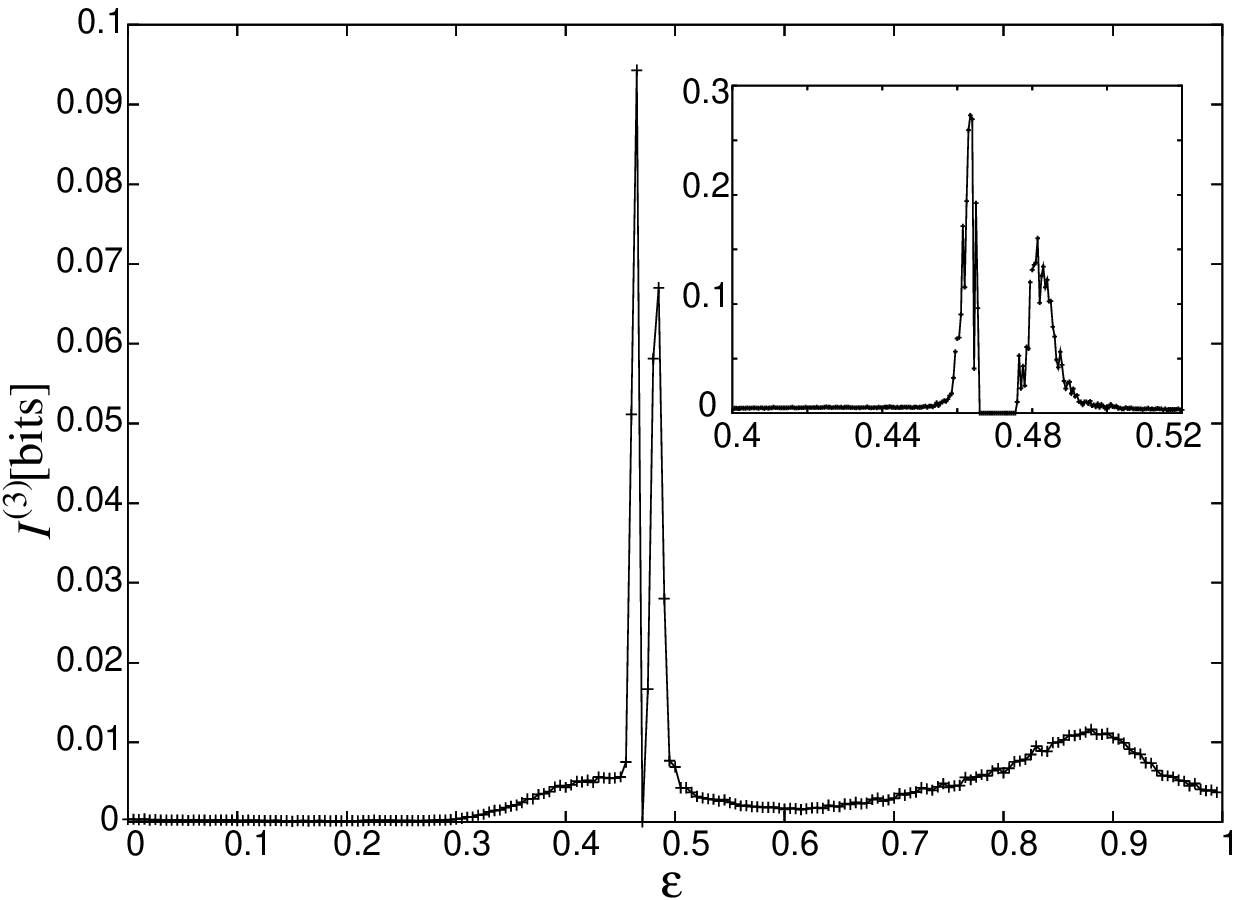}
  \includegraphics[height=6.35cm]{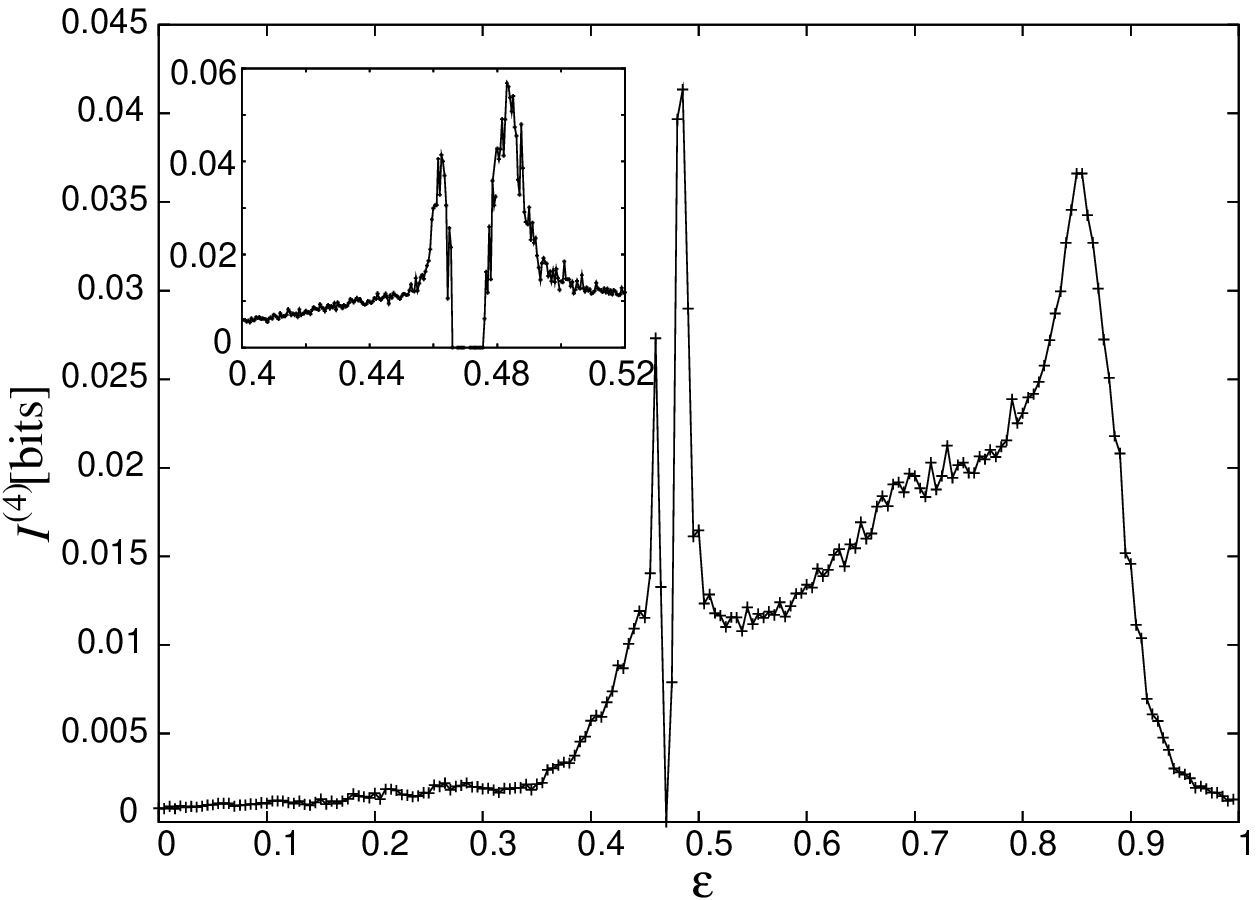}
  \includegraphics[height=6.35cm]{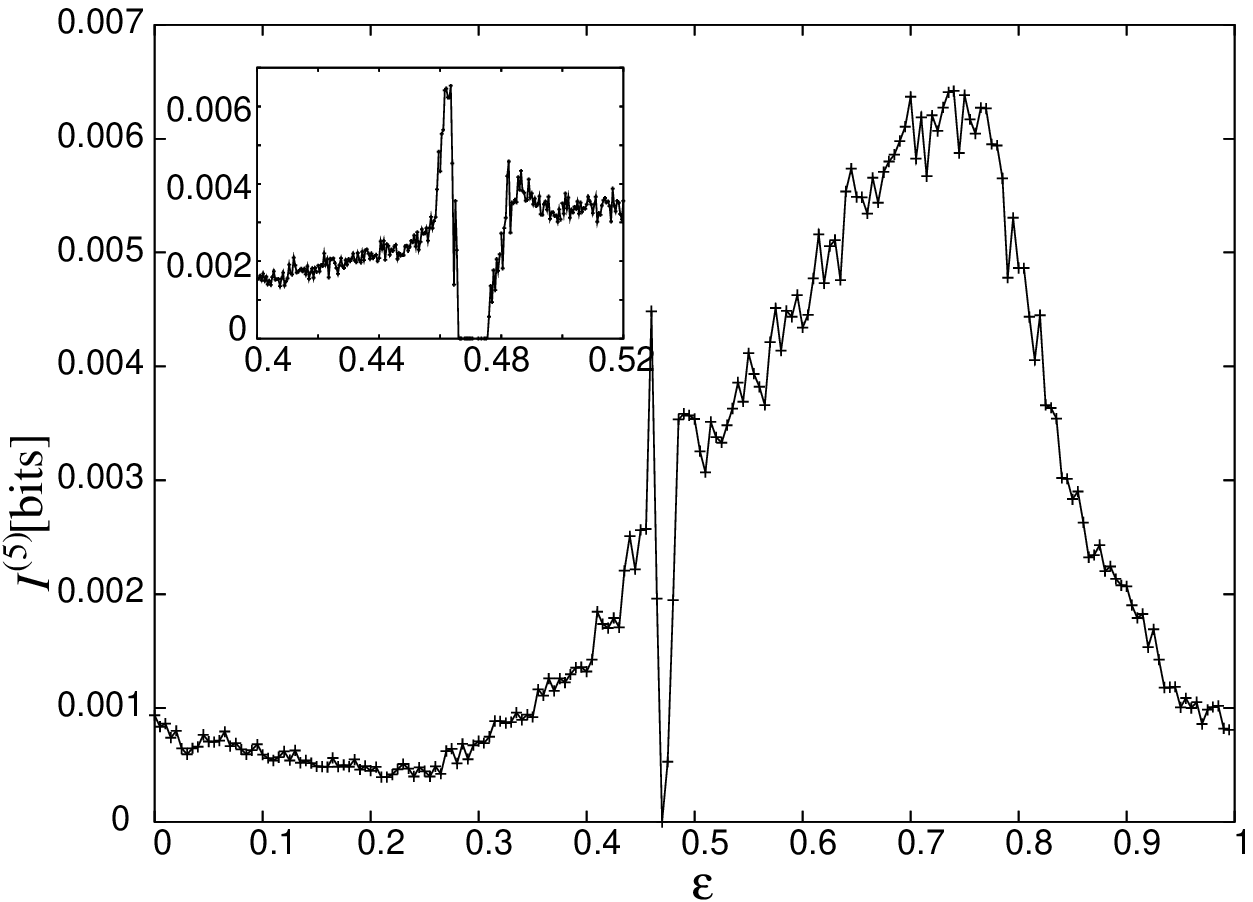}
  \includegraphics[height=6.35cm]{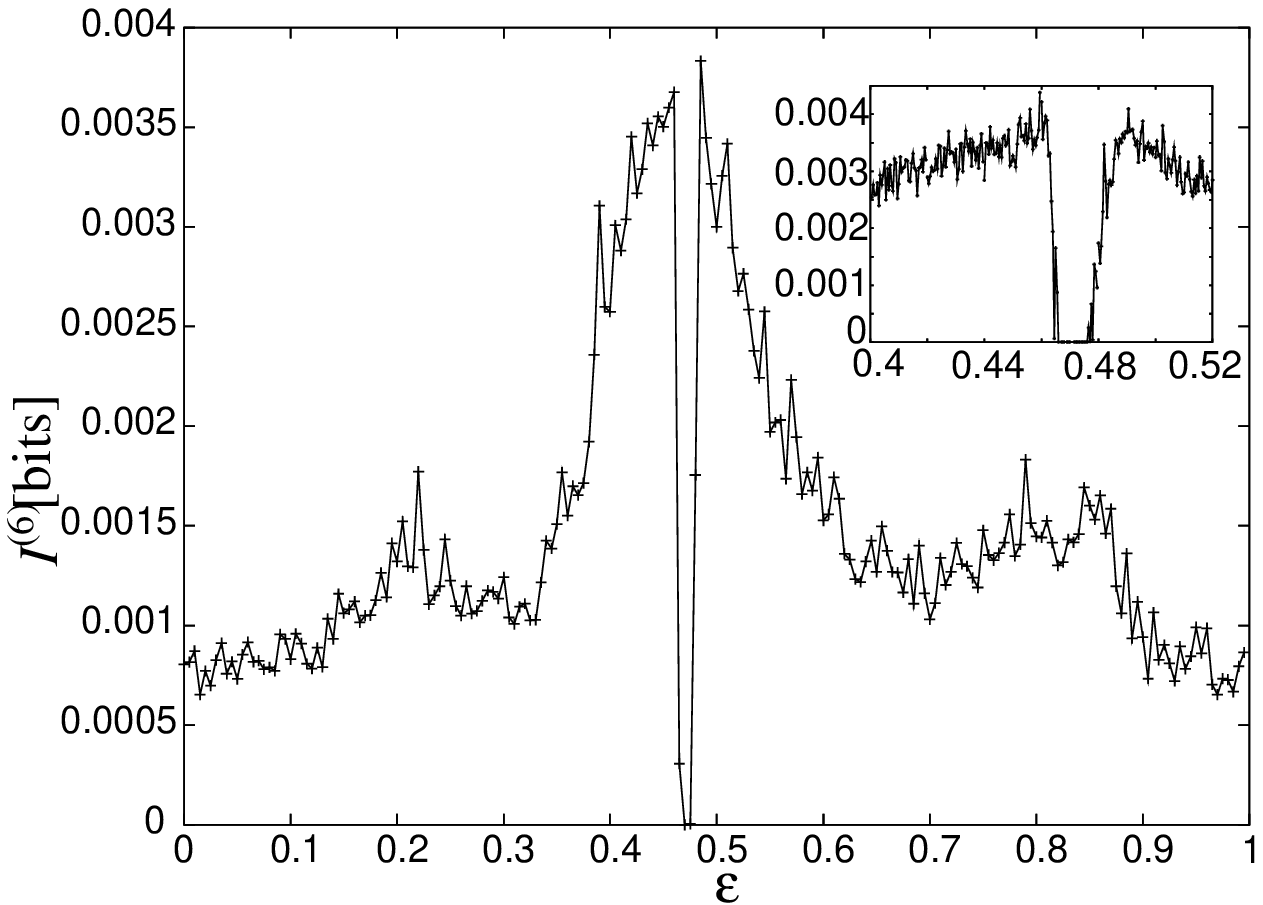}
  \caption{\label{fig:circle} The values of $I^{(k)}$ for the circle
    graph. The inlays show the parameter region $\epsilon \in [0.46,
    0.49]$. }
\end{figure*}

In Figures \ref{fig:full} and \ref{fig:circle}, the values of
$I^{(1)}$ to $I^{(6)}$ are plotted for the fully connected and the
circle graph, depending on the coupling $\epsilon$. For $k>6$, the
$I^{(k)}$ are very small and depend strongly on the random initial
condition, so that we don't evaluate them here. Figure \ref{fig:lyapu}
shows the behavior of the largest Lyapunov exponent for the two
network structures.  Different snapshots of the symbolic dynamics of
the circle graph are given in Figures \ref{fig:snapshotdynamicscirc},
\ref{fig:snapshotchaotic}, and \ref{fig:snapshotcomplex}.

For the fully connected graph, the nodes will be driven to
synchronized chaos if the coupling strength is high enough
\cite{kanekogcm90,jalanamritkar05}. We see this here as for $\epsilon
> 0.45$ $I$ is concentrated in $I^{(2)}$. It is interesting to observe
that the $I^{(k)}$ indicate complex dynamics taking place on the edge
of synchronized regimes. The largest Lyapunov exponent becomes
smallest at $\epsilon = 0.34$ (but not zero or negative, which would
indicate periodicity). The respective maxima of the $I^{(k)}$
correspond to this. They are located at $\epsilon_3 = 0.345$,
$\epsilon_{4}=0.42$, $\epsilon_{5}=0.355$, $\epsilon_{6}=0.35$. Also
note that $I^{(4)}$ has at least a local maximum at the parameter
value $0.35$.

\subsubsection{Circle Graph}

The circle graph of 10 nodes shows a specific ``nearly periodic''
behavior for parameter values $\epsilon \in [0.463, 0.481]$. This is a
specific feature due to the symmetry of the system. It is not present
for $N \neq 10$. We don't want to call the dynamics in that regime
complex, as there is a high degree of regularity. In the respective
parameter region, the average activity taken over all nodes is almost
constant. In the symbolic dynamics, 2 of the nodes are constant
themselves, 4 are periodic, while the remaining 4 are close to
periodicity(see Figure \ref{fig:snapshotdynamicscirc}). Obviously, the
complexity measures detect this phenomenon, as all $I^{(k)}, k\neq 2$
drop to zero.  More interestingly, when the parameter $\epsilon$ takes
values close to where the regularity occurs we can observe very high
values of the $I^{(k)}$ for $k \neq 2$. In this regime, the average
activity does not follow the one of the individual nodes, but
fluctuates on the same scale, instead of averaging out. This indicates
a complex dynamical structure driven by higher order correlations
among the nodes.  See Figure \ref{fig:snapshotcomplex} for a snapshot
of the dynamics. On the onset of the synchronization one can
additionally observe ``part time synchronization''.  This means that a
nearly periodic state, as described above, emerges, but is not
stable. After a couple of time steps (depending on the exact parameter
value), it dissolves again.

To understand the reason for the peak close to the periodic phenomenon
better we investigated the transition region closer. First, we took a
very long sequence that shows transitions. Then we separated the two
phases, saving them to different files and analyzed them
separately. The result here is, that the unordered state shows the
same $I$ vector as the region left of the peak, while the periodic
sequences of course have $I$ concentrated in $I^{\left( 2 \right)}$,
as the theory predicts. If the two types of sequences are mixed then
higher order correlations appear, leading to the peak. This
corresponds to the more general and unsolved problem whether the
complexity of a convex combination of two distributions is related to
the complexities of the individual constituents.

\section{Conclusion}
\label{sec:conclusion}
We have introduced complexity measures, suggested by
\cite{Amari01,jostaybertschiolbrich06}, that quantify the interactions
of $k$ parts of a system that cannot be explained by interaction
between fewer parts of that system. Then, after studying elementary
theoretical properties and their interpretation, we have explored the
measures in numerical studies. Using the elementary cellular automata
one can already study the behavior of the measures in special
cases. There are specific effects, such as emergence of higher
correlations, that can be found with our method. Commonly studied
quantities like correlation functions capture only part of what we
called $I^{(2)}$. Our measures therefore give a more global view.

Symbolic dynamics of coupled map lattices are far less understood from
a theoretical point of view. We have shown that using the symbolic
dynamics one can detect important dynamical properties of the
underlying real valued dynamics.  This is true despite the fact that
only spatial correlations are used in the computation of the
$I^{(k)}$.  We stress the point that our measures take as input only
the statistics of the system and make no model assumptions. They are
suitable for exploratory analysis, as it is done in sequence based
genetics. In an exploratory study one has no order parameters, as
these quantities need to be constructed from some model first.

Interestingly, we observed that complex behavior takes place on the
edge of synchronization, similar to the common wisdom of ``complex
dynamics on the edge of chaos''.  $I^{(2)}$ plays a special role in
this context. As synchronization phenomena are entirely captured by
pair interactions, we would suggest to measure complexity in terms of
$I^{(k)}$ for $k>2$. There is more work to do, investigating this
notion of complexity in different kinds of models.

Nevertheless the universality of the results remains to be
investigated, as our system size is very small and the computation of
the $I^{(k)}$ gets infeasible as the system grows. Therefore, for the
time being our approach is practically limited, while still
conceptually appealing.

\begin{acknowledgments}
  The authors wish to thank Sarika Jalan for discussions in the early
  stages of the work. This work is funded by the Volkswagen
  Foundation.
\end{acknowledgments}

\end{document}